\RequirePackage{fix-cm}
\documentclass[twocolumn]{svjour3}          
\smartqed  

\usepackage{cite}
\usepackage[pdftex]{graphicx}
\usepackage{ragged2e}
\usepackage[tight,footnotesize]{subfigure}
\usepackage{rotating}
\usepackage{graphicx}
\usepackage{amsmath,amssymb} 
\usepackage{array}
\usepackage{multirow}
\usepackage{colortbl}
\usepackage{amsfonts}
\usepackage{pifont}
\usepackage{xspace}
\usepackage{etoolbox}
\usepackage{overpic}
\usepackage{color}
\usepackage{microtype}

\usepackage{booktabs}
\usepackage{makecell}
\usepackage{threeparttable}
\usepackage{bbding}
\usepackage[numbers]{natbib} 
\usepackage{algorithm}
\usepackage{bm}
\usepackage{nicefrac}       
\usepackage{microtype}      
\usepackage{xcolor}         
\usepackage{appendix}
\usepackage{graphicx}
\usepackage{algorithm}
\usepackage{arydshln}
\usepackage{colortbl}
\usepackage{caption}  
\usepackage{graphicx} 
\usepackage{array}  
\usepackage{algpseudocode}

\newcommand{\orcid}[1]{\href{https://orcid.org/#1}{\includegraphics[width=10pt]{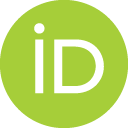}}}

\definecolor{myRed}{RGB}{219, 68, 55}
\definecolor{myGreen}{RGB}{15, 157, 88}
\definecolor{myBlue}{RGB}{66, 133, 244}

\def\etal{{\em et al.}}

\definecolor{mygray1}{gray}{.77}
\definecolor{mygray2}{gray}{.92}

\definecolor{CiteColor}{RGB}{0, 113, 188}

\usepackage[breaklinks=true,citecolor=blue, colorlinks]{hyperref}

\graphicspath{{./}}
\DeclareGraphicsExtensions{.pdf,.jpg,.png}

\usepackage{silence}
\definecolor{personcolor}{RGB}{234, 60, 49}
\definecolor{othervehiclecolor}{RGB}{96,81,242}
\definecolor{mygrey}{gray}{0.9}

\newcommand{\model}{DreamConnect}

\title{Connecting Dreams with Visual \\Brainstorming Instruction}

\author{
  Yasheng Sun$^{1}$\orcid{0000-0002-5245-7518} \and
  Bohan Li$^{2}$\orcid{0000-0002-5245-7518} \and
  Mingchen Zhuge$^{3}$\orcid{0000-0002-5245-7518}\\ 
  Deng-Ping Fan$^{4*}$\orcid{0000-0002-5245-7518} \and
  Salman Khan$^{5}$\orcid{0000-0002-5245-7518} \and \\
  Fahad Shahbaz Khan$^{5}$\orcid{0000-0002-5245-7518} \and
  Hideki Koike$^{1}$\orcid{0000-0002-5245-7518}
}

\authorrunning{Sun~\etal} 

\institute{
$^1$ CS, Tokyo Institute of Technology, Tokyo 152-8550, Japan. \\
$^2$ CS, Shanghai Jiaotong University, Shanghai 200240, China. \\
$^3$ Center of Excellence for Generative AI, KAUST, Thuwal 23955-6900, Saudi Arabia. \\
$^4$ CS, Nankai University, Tianjin 300350, China. \\
$^5$ CV Group, MBZUAI, Abu Dhabi 20015, UAE. \\
$^*$ Corresponding author: D.-P. Fan (dengpfan@gmail.com)\\
}

\date{Received: date / Accepted: date}

\journalname{}

\begin{document}

\maketitle

\begin{figure*}[h]
  \centering
  \includegraphics[width=1.0 \textwidth]{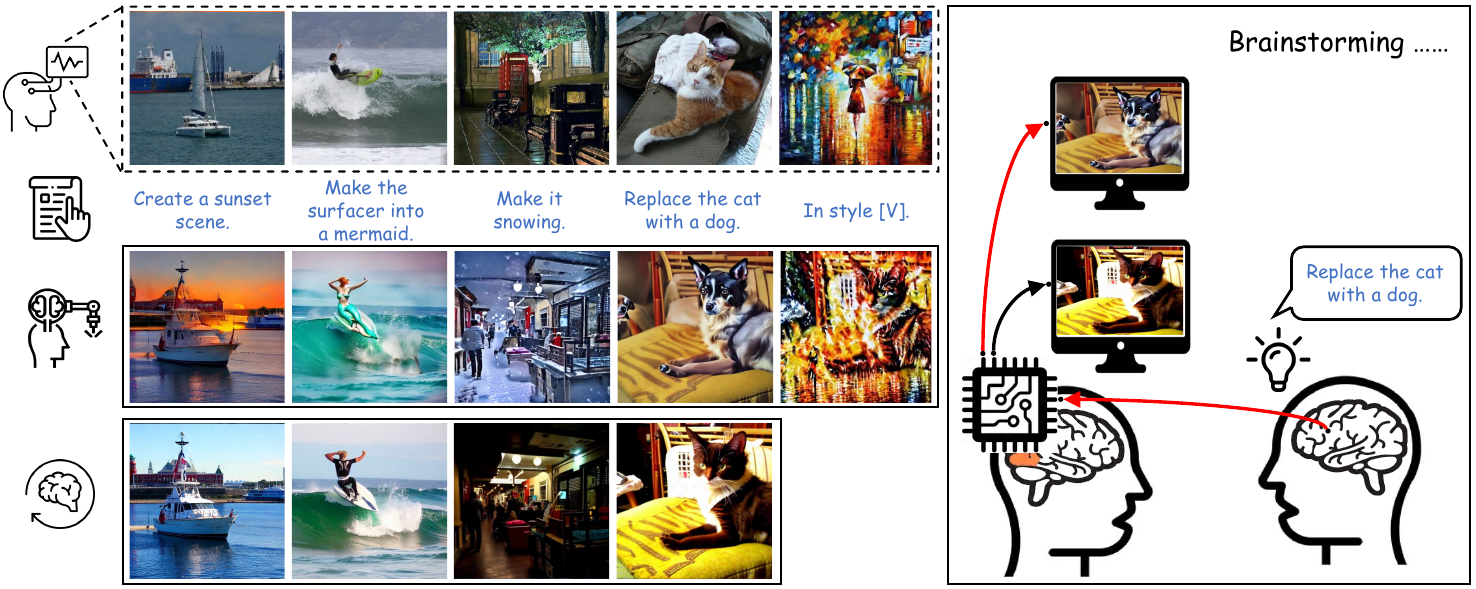} 

  \caption{\textit{Can dreams be connected and actively influenced in future applications?} As can be seen, \textbf{\model{}} precisely performs the desired operation on the visual content. For example, suppose someone imagines a lake view (see \textbf{the first row}) and another one considers changing it to a sunset scene (\textbf{the second row}). In that case, our system faithfully generates the desired sunset ambiance (\textbf{the third row}).}
  \label{fig:teaser}
\end{figure*}

\begin{abstract}

Recent breakthroughs in understanding the human brain have revealed its impressive ability to efficiently process and interpret human thoughts, opening up possibilities for intervening in brain signals.
In this paper, we aim to develop a straightforward framework that uses other modalities, such as natural language, to translate the original ``dreamland''.
We present \textbf{\model{}}, employing a dual-stream diffusion framework to manipulate visually stimulated brain signals. 
By integrating an asynchronous diffusion strategy, our framework establishes an effective interface with human ``dreams'', progressively refining their final imagery synthesis. 
Through extensive experiments, we demonstrate the method's ability to accurately instruct human brain signals with high fidelity. Our project will be publicly available on \textcolor{blue}{\href{https://github.com/Sys-Nexus/DreamConnect}{https://github.com/Sys-Nexus/DreamConnect}}.

\end{abstract}


\section{Introduction}
The emergence of deep generative models~\cite{rombach2022high,ho2020denoising,takagi2023high,chen2023seeing,scotti2023reconstructing,lu2023minddiffuser,sun2024contrast,shen2024neuro,xia2024dream} has effectively bridged the modality gap between functional Magnetic Resonance Imaging (fMRI)~\cite{ogawa1990oxygenation} and diverse signals like visual, language, and audio.
However, many studies~\cite{,bai2023dreamdiffusion,lan2023seeing,zeng2023controllable,vanrullen2019reconstructing,dado2022hyperrealistic,shen2019deep,seeliger2018generative,NEURIPS2022_bee5125b,gu2022decoding,liu2024mind} focus on recovering the stimulated signal from brain activity, while interaction with reality requires active communication with human brain signals.

In this work, we explore a novel setting as shown in Fig.~\ref{fig:teaser} -- \emph{Can ``dreams'' be connected and actively influenced?} Such capability will present a novel and convenient interaction paradigm among people, facilitating numerous applications such as creative design via brainstorming.
The essence of this capability lies in steering the human ``dreams'' towards desired directions, ultimately \emph{enabling concept manipulation through direct operation on the fMRI signal}. 
This paper mainly focuses on human ``dreams'' as a visual modality, but the fMRI signal can also store other types of stimulated information, such as audio~\cite{defossez2023decoding} or language~\cite{zhao2024mapguide}. Consequently, our proposed approach can be seamlessly extended to these scenarios.

This task requires direct instruction of fMRI signals, which presents notable challenges on two fronts: 1) The inherent abstractness and ambiguity of collected brain signals often impede the comprehension of their content; 
2) A significant modality gap exists between fMRI signals and natural language instructions, hindering the learning of accurate associations between brain activity features and language instructions. Since not all the information within the fMRI signals is relevant to the intended instruction, successful instruction requires \emph{pinpointing and modifying the relevant features while preserving the irrelevant ones}.  

We introduce \textbf{\model{}}, a dual-stream diffusion framework tailored for fMRI signal instruction to address these challenges. The key is to \emph{effectively guide language instructions to hone in on and modify the pertinent information embedded within fMRI signals}. To establish a correlation between these distinct signals, we first employ a Stable Diffusion (SD)~\cite{rombach2022high} backbone to interpret brain activity using visual prior knowledge. Then, we integrate language instructions into a parallel SD backbone to manipulate the acquired visual priors.  
This parallel forwarding approach has proven to be effective in improving network learning ability by leveraging the homogeneous nature of the extracted features~\cite{zhang2023adding}. To bridge these two streams, i.e., the interpretation stream and instruction stream, we introduce an adaptor network designed to modulate the intermediate visual contents from the interpretation stream hierarchically.

Ideally, with this design, the framework will operate under a \emph{progressive interpretation and instruction paradigm, concurrently identifying and manipulating correlated regions}. However, the diffusion operation behaves distinctively along the temporal dimension, synthesizing varied content across different time steps~\cite{zhang2023prospect,Zhang_2023_inst}. At specific time steps, the second stream must identify specific visual features for instruction, necessitating that the first stream completes their formation beforehand. Therefore, we devise an asynchronous diffusion strategy to ensure that the first stream has adequate time to develop overall semantics, thereby facilitating smooth manipulation progress. Additionally, to further guide the model toward intended spatial locations, we introduce a Large Language Model (LLM)~\cite{wei2022emergent} agent tasked with identifying regions relevant to instructions. Using the identified areas, we implement a region-aware attention strategy to ensure that the instructions are applied precisely within those regions.
\textbf{Our main contributions can be summarized as:}

\noindent \textbf{(1)} We propose a visual brainstorming instruction system, named \textbf{\model{}}, empowering users to effectively interfacing with human ``dreams'' to perform desired operation.

\noindent \textbf{(2)} We devise a dual-stream diffusion network coupled with an adaptor to seamlessly translate fMRI signals towards their intended directions.

 \noindent \textbf{(3)} We introduce an asynchronous diffusion strategy, enhanced with LLM-guided region-aware manipulation, to facilitate faithful instruction within pertinent regions from both temporal and spatial perspectives.
\noindent \textbf{(4)} While designed for language-guided instruction, our framework can readily adapt to visual instruction or multi-modal instruction with minimal adjustments.


\section{Related Work}

\noindent \textbf{Brain Activity Recognition.} 
Earlier studies~\cite{vanrullen2019reconstructing,dado2022hyperrealistic,shen2019deep,seeliger2018generative,NEURIPS2022_bee5125b,gu2022decoding,ferrante2022semantic} aimed to reconstruct visual stimuli by decoding brain signals into the latent space of Generative Adversarial Networks~\cite{goodfellow2020generative}. Given the significant capabilities of diffusion models in image generation~\cite{rombach2022high,ho2020denoising}, numerous investigations~\cite{takagi2023high,chen2023seeing,scotti2023reconstructing,lu2023minddiffuser,bai2023dreamdiffusion,lan2023seeing,zeng2023controllable,li2024neuraldiffuser,fang2024alleviating,li2024visual} have explored the use of their image prior for high-quality reconstruction. To improve reconstruction performance, some studies~\cite{zeng2023controllable,lu2023minddiffuser} exploit low-level signals within fMRI to achieve spatially consistent predictions by imposing spatial constraints during the generation process. Another set of studies~\cite{liu2023brainclip} has sought to leverage contrastive learning, such as CLIP, to improve semantic alignment in reconstruction. Certain works demonstrate the effectiveness of masked pretraining~\cite{he2022masked,bai2023dreamdiffusion} of brain signals on large datasets, facilitating robust latent space learning. 
Moreover, novel applications related to brain activity~\cite{sun2023tuning,sun2024neurocine,wang2024mindbridge,nikolaus2024modality,xia2024umbrae,chen2024bridging,huo2024neuropictor} have emerged. For instance, Mind-Video~\cite{chen2023cinematic} extends image reconstruction to video reconstruction, while UniBrain~\cite{mai2023unibrain} not only reconstructs visual signals but also generates corresponding captions.

\paragraph{Image Manipulation.} 
With a plethora of diverse generative priors embedded within powerful diffusion architectures, numerous studies~\cite{avrahami2023blended,couairon2022diffedit,meng2021sdedit,nichol2021glide,sun2023imagebrush,sun2024avi} have endeavored to adapt them for image manipulation. To facilitate the editing of desired semantic regions within spatial feature maps, researchers have proposed various techniques such as cross-attention injection~\cite{hertz2022prompt,balaji2022ediffi} and introduced semantic loss~\cite{tumanyan2023plug} to constrain spatial features during the generation process. Additionally, to provide explicit guidance towards intended instructions, subsequent works~\cite{brooks2023instructpix2pix,geng2023instructdiffusion} leverage language-based instructions to fine-tune the SD model, showcasing impressive capabilities in conforming to desired editing directions.
However, these approaches primarily operate on raw images. Few have delved into manipulating stored visual content with another modality, which holds potential for diverse applications such as fostering friendly human-computer interaction.

\paragraph{LLM Agent For Visual Assistance.}
Large language models (LLMs) have showcased profound capabilities as remarkable reasoning engines in various tasks~\citep{zhuge2023mindstorms,wei2022chain,huang2022towards,Touvron2023LLaMAOA,brown2020language}, due to their emerging ability~\citep{wei2022emergent}.
With rich contextual prior knowledge, LLMs adeptly tackle a multitude of visual-language tasks~\citep{zhuge2024language,alayrac2022flamingo,zhang2023multimodal,chakrabarty2023spy} via appropriate prompting adaptation.
Through visual instruction tuning, LLMs can discern image content, reason about involved events, and generate plausible responses~\citep{li2023blip,liu2024visual,zhu2023minigpt}. Subsequent studies~\citep{koh2024generating,sun2023generative,wu2023visual} have shown that LLMs can naturally provide visual feedback by leveraging an image rendering backbone such as a diffusion model.
Recently, researchers have employed LLMs to facilitate the image generation process~\citep{NEURIPS2023_3a7f9e48,fu2023guiding}, where the language model exhibits impressive capabilities in layout reasoning and instruction execution.

\section{Methodology}

\begin{figure*}[t!]
  \centering
   \includegraphics[width=\linewidth]{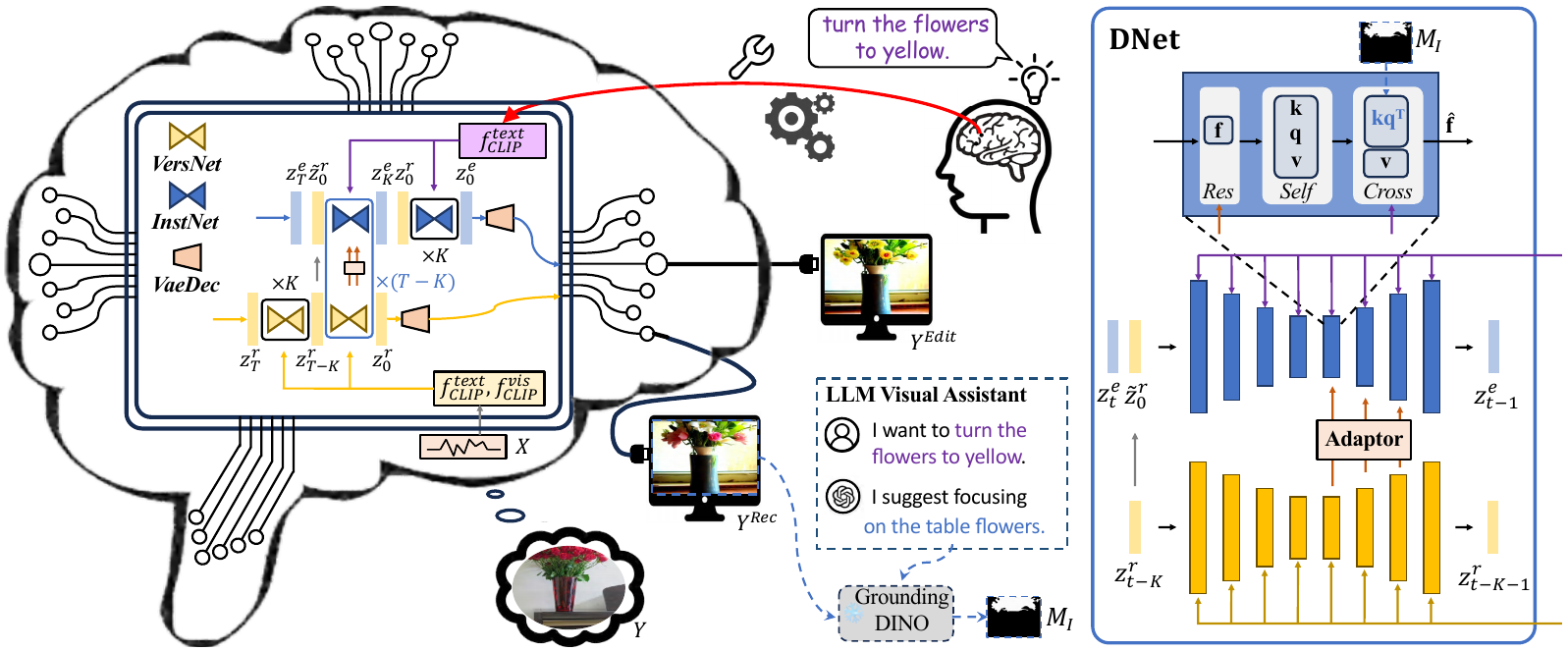}
  \caption{\textbf{Illustration of the proposed \textbf{\model{}} framework.} 
    A person's biological signal indicated by fMRI sequences $X$, will be activated within his brain according to his ``dreams'' represented by the visual stimuli $Y$. 
    Our system targets to interface with $X$ via a natural language instruction $I$. 
    Specifically, the fMRI signal is regressed to CLIP text and visual embedding, $f_{\text{CLIP}}^{text}$ and $f_{\text{CLIP}}^{vis}$, which are leveraged to aligned with visual content $z^r_t$ in \emph{\textbf{VersNet}}~\cite{xu2023versatile}. After modulated by an adaptor, its intermediate spatial features are fed to \emph{\textbf{InstNet}}~\cite{geng2023instructdiffusion}. Encoded by CLIP text encoder, the human instruction $I$ is injected to \emph{\textbf{InstNet}} to modulate these features toward intended direction. 
  }
  \label{fig:arch}
\end{figure*}

In this section, we discuss the details of the proposed \emph{Visual Brainstorming Instruction System}, termed \textbf{\model{}}. The primary objective of this framework is to develop a model capable of providing instruction in brain activity. To achieve this, the model must possess the ability to associate human instructions with fMRI signal and modify its embedded pertinent visual information. The overall pipeline is depicted in Fig.~\ref{fig:arch} where we devise an asynchronous dual-stream diffusion architecture coupled with an adaptor to promote \emph{progressive interpretation and instruction}. To facilitate faithful instruction within pertinent regions from both temporal and spatial perspectives, we introduce an asynchronous diffusion strategy and an LLM-guided region-aware manipulation operation.

\subsection{Dual Stream Diffusion Instruction}
\textbf{Problem Formulation.} Given a visual stimuli $Y \in \mathcal{R}^{3 \times H \times W}$, a person's biological signal revealed by fMRI, $X \in \mathcal{R}^{N_f}$, will be activated, where ${N_f}$ represents the number of relevant voxels. The traditional image reconstruction task targets to recover its indicted visual content $Y^{Rec}$ from $X$. In contrast, we study directly manipulating the entailed information to $Y^{Edit}$ following a piece of natural language instruction $I$. 
Accomplishing this objective requires a comprehensive cross-modal association between the fMRI signal and the specified instruction. To resolve this issue, we leverage the rich contextual prior within StableDiffusion (SD) to bridge the modality gap due to its effective exploitation in numerous visual tasks. The key is to derive a unified diffusion framework to tackle feature alignment and interaction among distinct modalities progressively.

\paragraph{Preliminary on Diffusion Models.} 
Diffusion models~\cite{ho2020denoising} stand out due to its stable training objective and exceptional ability to generate high-quality images. It operates by iteratively denoising Gaussian noise to produce the image $\bm{x}_0$.
Typically, the diffusion model assumes a Markov process~\cite{rabiner1989tutorial} wherein Gaussian noises are gradually added to a clean image $\bm{x}_0$ based on the following equation:
\begin{equation}
    \bm{x}_t = \sqrt{\alpha_t} {\bm{x}}_{0} + \sqrt{1-\alpha_t} \bm{\epsilon} ,
\label{eq:forward}
\end{equation}
where $\bm{\epsilon}\!\sim\!\mathcal{N}(0, \mathbf{I})$ represents the additive Gaussian noise, $t$ denotes the time step and $\alpha_t$ is scalar functions of $t$.
Our objective is to devise
a neural network $\bm{\epsilon}_\theta(\bm{x}_t, t, c)$ to predict the added noise $\bm{\epsilon}$. 
Empirically, a simple mean-squared error is leveraged as the loss function:
\begin{equation}
    L_{simple} := \mathbb{E}_{\bm{\epsilon} \sim \mathcal{N}(0, \mathbf{I}), \bm{x}_0, c}\Big[ \Vert \bm{\epsilon} - \bm{\epsilon}_\theta(\bm{x}_{t}, c) \Vert_{2}^{2}\Big] \, ,
    \label{eq:LDM_loss}
\end{equation}
where $\theta$ represents the learnable parameters of our diffusion model, and $c$ denotes the conditional input to the model. 
To improve the computational efficiency, latent diffusion models (LDM)~\cite{rombach2022high} proposes to operate in a lower-dimensional latent space that encodes $\bm{x}_t$ to $\bm{z}_t$ through VAE~\cite{esser2021taming}.

\paragraph{Dual Stream Diffusion Network.} The overall synthesis process is visualized in Fig.~\ref{fig:arch}. Given a sequence of fMRI signal $X \in \mathcal{R}^{N_f}$ and a sentence of instruction $I$ such as \emph{turn the flowers to yellow}, our objective is to translate $X$ to $Y^{Edit}$ according to the provided instruction.
The $N_f$ denotes the feature length of collected fMRI signal. 
To connect instruction description with fMRI signal, we devise a dual-stream diffusion network for progressive \emph{interpretation and instruction} as shown in the right side of Fig.~\ref{fig:arch}.
Accepting information from fMRI signal, the bottom stream aims to hallucinate visually appealing and semantically consistent content for faithful interpretation of the brain signal.
On top of this stream, another stream  aims to manipulate its decoded visual content according to user-specified instructions. Conditioned on brain signal $X$ and instruction $I$, the dual-stream diffusion network denoises the VAE latent codes $\textbf{z}_t^r$ and $\textbf{z}_t^e$ at each time step $t$. Formally,
\begin{equation}
    [\textbf{z}_{(t-1)}^r, \textbf{z}_{(t-1)}^e] = \textbf{\text{DNet}}_{t}([\textbf{z}_t^r, \textbf{z}_t^e]; I, X).
\end{equation}

For the purpose of bridging visual content and brain activity signal, we choose VersatileDiffusion (VD)~\cite{xu2023versatile} as the bottom-stream backbone because both image and text condition are taken into consideration following typical fMRI-based reconstruction works~\cite{ozcelik2023brain,scotti2023reconstructing}.
Specifically, the UNet of VD adaptively exploit extracted features from CLIP image and text encoder through cross attention.
We leverage their pretrained parameters to initialize the UNet, thereby keeping most generative image prior.
Then the UNet is freezed and we devise a mixed mapping strategy to predict both CLIP text and visual embed, $f_{\text{CLIP}}^{text}$ and $f_{\text{CLIP}}^{vis}$, within a diffusion paradigm.
The mapping network architecture and training strategy adopts align-before-predict paradigm with previous work~\cite{ramesh2022hierarchical,scotti2023reconstructing} but we operates on both visual and text space that co-manipulates VD in a mixed manner.

With the interpreted visual content, upper stream targets to guide them towards the desired instruction direction. A natural way to address this is to inject the CLIP text embedding of the instruction $I$ via cross attention, which is leveraged to modulate the spatial features within a diffusion UNet backbone. To connect these two streams, an adaptor network is introduced to adjust the intermediate features of interpreted visual content for better instruction. The adjusted features are injected only in the decoder part of the UNet due to its clear layout and semantic feature construction~\cite{zhang2023adding}. 
To maintain high consistency over the generated shape and layout, we propose to modify the features within residual blocks~\cite{tumanyan2023plug}. To further enhance control over the aligned visual content, we also incorporate the obtained VAE latent $\textbf{z}_t^r$ from bottom stream as concatenation of our input. Note that we convert $\textbf{z}_t^r$ at time $t$ to its denoised estimation $\tilde{\textbf{z}}_0^r$ at time step $t=0$ for consistent representation. Formally,
\begin{equation}
    \tilde{\textbf{z}}_0^r = (\textbf{z}_t^r - \sqrt{1-\overline{\alpha}_t} \bm{g}_{\theta} (\textbf{z}_t^r)) / \sqrt{\overline{\alpha}}_t,
\end{equation}
where the $\overline{\alpha}_t := \prod_{s=1}^t \alpha_s$ increases along with $t$ and $\bm{g}_{\theta}$ indicates the network of bottom stream.
For this stream, we initialize it with pretrained UNet from InstructDiffusion~\cite{geng2023instructdiffusion}.

\subsection{Faithful Instruction via Feature Exploitation}
The core of our framework lies on bridging fMRI signal and language instruction with visual modality. Thus fully utilization of the visual information plays a crucial role in successful instruction. We aim to make most of them from both temporal and spatial perspectives to facilitate faithful instruction.
\paragraph{Asynchronous Diffusion Strategy.} One issue of above strategy is that the aligned visual feature $\textbf{z}_t^r$ might not well prepared for manipulation at time step $t$. This problem becomes more severe especially at early stages because its semantic feature is not well formed, which brings difficulty for the instruction to identify pertinent areas to be edited. To address this, we propose an asynchronous approach where the instruction stream is enforced to lag behind the first stream for $K$ steps. Such practice will offer the first stream sufficient time to develop overall layout or shape semantics. Therefore, the improved version of diffusion process can be written as
\begin{equation}
    [\textbf{z}_{(t\textcolor{red}{-K}-1)}^r, \textbf{z}_{(t-1)}^e] = \textbf{\text{DNet}}_{t}([\textbf{z}_{t\textcolor{red}{-K}}^r, \textbf{z}_t^e]; I, X).
\end{equation}
The $K$ represents the number of time steps that the instruction stream lags behind, which we empirically adopt $K=15$ in our experiment.

\paragraph{LLM-Guided Region-Aware Instruction.} The asynchronous strategy implicitly provides higher chances for user instructions to operate on the appropriate regions. This inspires us to seek mechanisms to explicitly restrict feature manipulation within desired regions. Given the robust reasoning capabilities of LLMs, we leverage them as agents to aid in specifying the intended editing region. As depicted in lower part of Fig.~\ref{fig:arch}, we provide the instruction information to the language model and inquiry for the locations that requires focusing on. 
With the obtained advice, we employ it as an text prompt to guide the \emph{Grounding DINO} model to identify the pertinent spatial areas $M^I$ of the reconstructed image $Y^{Rec}$. Then the mask $M_I$ is used to restrict cross-attention. For each cross-attention layer, the instruction features are projected into context values $\textbf{v}$ and keys $\textbf{k}$ while visual features are projected into queries $\textbf{q}$. Its output of this block is given by 
\begin{equation}
    \hat{\textbf{f}} = \textbf{A} \textbf{v} \text{  where  } \textbf{A} = \text{Softmax}(\textbf{q}\textbf{k}^T).
\end{equation}
Intuitively, the attention map $\textbf{A}$ determines distribution ratio of context features. 
Here the instruction irrelevant region is expected to be untouched. A natural way is to mask out those features outside of $M_I$ by replacing them with attention maps activated by null instruction. Formally,
\begin{equation}
    \textbf{A}_M = \textbf{A}_I M_I + \textbf{A}_{null} (1-M_I)
\end{equation}
where the updated attention map $\textbf{A}_M$ is a masked combination of attention maps activated by instructions and null instructions, $\textbf{A}_I$ and $\textbf{A}_{null}$, respectively. The guiding mask $M_I$ is interpolated to the spatial size of each UNet block. 


\section{Experiments}
\begin{figure*}[t!]
  \centering
  \includegraphics[width=.98\linewidth]{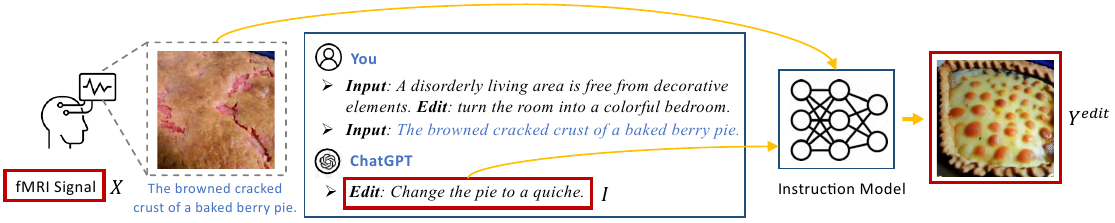}
  \caption{\textbf{Illustration of Dataset Pipeline.} 
  Given the paired fMRI signal $X$ and its corresponding visual stimuli $Y$ in NSD~\cite{allen2022massive}, we first query LLMs with image caption to obtain possible instruction $I$. Then the instruction, coupled with visual stimuli, is utilized prompt a visual instruction model for manipulated image $Y^{edit}$.
  Finally, the triplets of $(X, I, Y^{edit})$, are constructed (red boxes).
  }
  \label{fig:data_pipeline}
\end{figure*}

\textbf{Datasets.} To validate our proposed approach, we leverage the NSD~\cite{allen2022massive} dataset, the largest neuro-imaging dataset containing densely sampled fMRI data. 
Participants are asked to view 9000-10000 different pictures over 30-40 MRI scanning sessions. 
These images are extracted from COCO dataset~\cite{lin2014microsoft}, and corresponding captions are also available. In our experiments, we choose subject 1 from NSD.
For each subject, a total of 9841 image stimuli are presented, with 982 images set aside for validation. 
In our setting, we require triplets of fMRI data, instructions and edited images to train and evaluate our model.
Since such a dataset does not exist, we augment the NSD dataset using Large Language Models (LLMs), as illustrated in Fig.~\ref{fig:data_pipeline}. Specifically, LLMs generate instructions, while a pre-trained visual instruction model accounts for obtaining corresponding edited image. For easy access, we utilize text-davinci-003 engine of ChatGPT~\cite{brown2020language} for in-context instruction synthesis and instructDiffusion~\cite{geng2023instructdiffusion} for visual instruction. Replacing these with more advanced models such as GPT-4V(ison)~\cite{openai2023gpt} or DALL·E 3~\cite{openai2023dalle3} could further improve the dataset quality. 

\paragraph{Implementation. } During the training stage, we freeze the UNet backbones of both the first and second streams to fully leverage the visual priors pre-trained. The overall training process is divided into two stages. In the first stage, we target to translate fMRI signal into the visual domain by regressing the intermediate features of CLIP vision/text embedding. The architecture of the regression network and the training paradigms follow protocols similar to MindEye~\cite{scotti2023reconstructing}. Once the parameters of this stream have been trained, they are frozen. The adaptor that bridges two streams is finetuned using our extended synthesized dataset. We employ the Mean Squared Error (MSE) Loss~\cite{ho2020denoising} typical in diffusion models as our optimization objective.
Since each image stimulus includes three trials, a total of 24980 trials and 8859 visual stimuli are involved in the training set. For fMRI data with multiple trials, we average their responses for consistency. 
For our framework and all the compared models, we set the classifier-free scale to 7.5 for image instruction.
Our models are implemented in PyTorch~\cite{paszke2019pytorch} and trained using 80G Tesla A100 GPUs.

\paragraph{Comparisons.} Although our framework is designed for fMRI signal instruction, it also supports basic function of reconstruction. Therefore, we validate our method from both the reconstruction and the instruction perspectives. For image reconstruction, we compare our method with the best models currently available that support fMRI-based image reconstruction: Takagi \emph{et.al.}~\cite{takagi2023high}, UniBrain~\cite{mai2023unibrain}, Brain-Diffuser~\cite{ozcelik2023brain} and MindEye~\cite{scotti2023reconstructing}. 
Takagi \emph{et.al.}~\cite{takagi2023high} pioneered the utilization of image priors from Stable Diffusion (SD)~\cite{esser2021taming} for fMRI-based image reconstruction. 
UniBrain~\cite{mai2023unibrain} demonstrates the capability to translate fMRI signals to both image captions and images simultaneously. 
Brain-Diffuser~\cite{ozcelik2023brain} and MindEye~\cite{scotti2023reconstructing} showcase the effectiveness of mapping to CLIP space in the VersatileDiffusion~\cite{xu2023versatile} setting.
For instruction, we compare our method with state-of-the-art image manipulation approaches, including SDEdit~\cite{meng2021sdedit}, InstructPix2Pix~\cite{brooks2023instructpix2pix}, InstructDiffusion~\cite{geng2023instructdiffusion} and MagicBrush~\cite{zhang2024magicbrush}. 
SDEdit~\cite{meng2021sdedit} proposes to edit image with Stochastic Differential Equation (SDE).
InstructPix2Pix~\cite{brooks2023instructpix2pix} directly trains a Stable Diffusion model conditioned on instructions. InstructDiffusion~\cite{geng2023instructdiffusion} and MagicBrush~\cite{zhang2024magicbrush} further extend the dataset and improve the model performance for instruction-based image editing.

\subsection{Quantitative Evaluation}



\begin{table*}[t]
\setlength{\tabcolsep}{9pt} 
\renewcommand{\arraystretch}{1.1} 
\centering
\caption{\textbf{Quantitative results of image reconstruction on the Natural Scenes Dataset (NSD)~\cite{allen2022massive}.}}
\label{table:rec}
\begin{tabular}{lcccccccc}
\toprule
 & \multicolumn{4}{c}{Low-Level Metrics} & \multicolumn{4}{c}{High-Level Metrics} \\
\cmidrule(lr){2-5} \cmidrule(lr){6-9}
Method & PixCorr $\uparrow$ & SSIM $\uparrow$ & Alex(2) $\uparrow$ & Alex(5) $\uparrow$ & Incep $\uparrow$ & CLIP $\uparrow$ & Eff $\downarrow$ & SwAV $\downarrow$ \\
\midrule
Takagi \emph{et.al.}~\cite{takagi2023high} & - & - & 0.830 & 0.830 & 0.760 & 0.770 & - & - \\
UniBrain~\cite{mai2023unibrain} & 0.249 & \underline{0.330} & 0.929 & 0.956 & 0.878 & 0.923 & 0.766 & 0.407 \\
Brain-Diffuser~\cite{ozcelik2023brain} & 0.254 & \textbf{0.356} & 0.942 & 0.962 & 0.872 & 0.915 & 0.775 & 0.423 \\
MindEye~\cite{scotti2023reconstructing} & \underline{0.309} & 0.323 & \underline{0.947} & \underline{0.978} & \underline{0.938} & \textbf{0.941} & \textbf{0.645} & \underline{0.367} \\
\hline
\rowcolor{mygrey} \textbf{\model{}} & \textbf{0.327} &  0.315 & \textbf{0.951} & \textbf{0.978} & \textbf{0.939} & \underline{0.934} & \underline{0.653} & \textbf{0.360} \\
\bottomrule
\end{tabular}
\end{table*}



\begin{table*}[t]
\setlength{\tabcolsep}{11pt} 
\renewcommand{\arraystretch}{1.1} 
\centering
\caption{\textbf{Quantitative results of image instruction on the Natural Scenes Dataset (NSD)~\cite{allen2022massive}.}}
\label{table:edit}
\begin{tabular}{lccccc}
\toprule
Method & InstructPix2Pix~\cite{brooks2023instructpix2pix} & InstructDiff~\cite{geng2023instructdiffusion} & MagicBrush~\cite{zhang2024magicbrush} & SDEdit~\cite{meng2021sdedit} & \textbf{\model{}} \\
\midrule
\textbf{CLIP-I} & \textbf{0.664} & 0.643 & 0.649 & 0.577 & \cellcolor{mygrey} \underline{0.657} \\
\textbf{DiNO-I} & \textbf{0.305} & 0.274 & 0.289 & 0.181 & \cellcolor{mygrey} \underline{0.301} \\
\textbf{CLIP-D} & 0.090 & \underline{0.113} & 0.105 & 0.101 & \cellcolor{mygrey} \textbf{0.114} \\
\bottomrule
\end{tabular}
\end{table*}


\begin{table*}[t!]
\setlength{\tabcolsep}{22pt} 
\renewcommand{\arraystretch}{1.1} 
\centering
\caption{\textbf{The ablation over model design on Natural Scenes Dataset (NSD)~\cite{allen2022massive}.}}
\label{table:ablation}
\begin{tabular}{lcccc}
\toprule
Method & wo / Asynch & wo / Adaptor Injection & wo / LLMs & \textbf{Full Model} \\
\midrule
\textbf{CLIP-I} & 0.639 & 0.600 & 0.650 & \cellcolor{mygrey} \textbf{0.657} \\
\textbf{Dino-I} & 0.299 & 0.184 & 0.292 & \cellcolor{mygrey} \textbf{0.301} \\
\textbf{CLIP-D} & 0.112 & \textbf{0.135} & 0.118 & \cellcolor{mygrey} 0.114 \\
\bottomrule
\end{tabular}
\end{table*}


\textbf{Evaluation Metric.}  For image reconstruction, we use both low-level and high-level metrics to measure the semantic correctness of our results. Specifically, we choose \textbf{PixCorr}, \textbf{SSIM}, \textbf{Alex(2)} and \textbf{Alex(5)} for low-level measurements while \textbf{Incep}, \textbf{CLIP}, \textbf{Eff} and \textbf{SwAV} are utilized for high-level semantic evaluation, following the evaluation protocol of MindEye~\cite{scotti2023reconstructing}.
For instruction-based evaluation, we use CLIP~\cite{radford2021learning} image similarity (\textbf{CLIP-I}) and Dino-V2~\cite{oquab2023dinov2} image similarity (\textbf{Dino-I}) to measure the cosine similarity between edited and original images. Additionally, we use CLIP text-image direction similarity~\cite{gal2022stylegan} (\textbf{CLIP-D}) evaluates how changes in images correspond to changes in their captions.

\paragraph{Evaluation Results.} The comparison results for image reconstruction from fMRI signals are presented in Table~\ref{table:rec}. Our results exhibit superior performance in low-level semantic consistency and competitive results in high-level metrics. One possible reason for this is the mixed mapping strategy in the first stream, where we regress both CLIP visual and text embeddings via latent diffusion. We speculate that this practice helps strike a balance between low- and high-level semantic formation. Notably, although our framework is not specifically designed for the reconstruction task, it achieves performance comparable to specialized approaches.

For image instruction performance, we present the comparison results in Table~\ref{table:edit}. Our method achieves superior results in Dino-V2 image similarity and CLIP text-image direction (CLIP-D) similarity, demonstrating its capability to manipulate visual imagery underlying fMRI signals aligned with human instruction. However, in the CLIP image similarity metric, our approach scores lower than InstructPix2Pix~\cite{brooks2023instructpix2pix} because they tend towards under-editing, making minimal changes to the input images. Consequently, their results show inferior performance in terms of the CLIP-D metric.

\subsection{Qualitative Evaluation}
\begin{figure*}[t!]
  \centering
  \includegraphics[width=\linewidth]{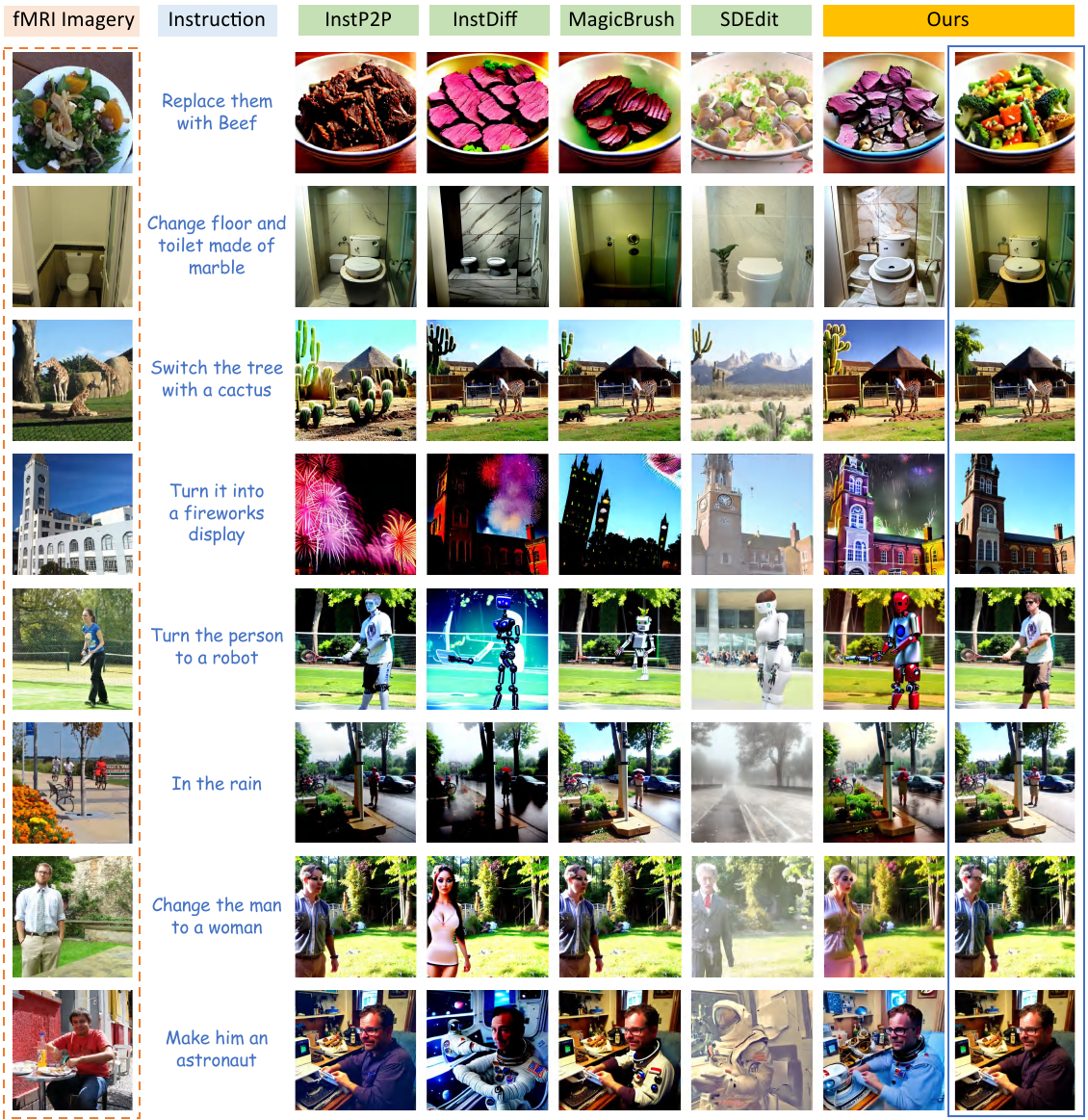}
  \caption{\textbf{Qualitative Comparison.} In the first column list the visual stimulus of fMRI signal while the last column demonstrates input images used by other approaches. InstP2P~\cite{brooks2023instructpix2pix} struggles on capturing instruction intention (see last 4 rows). InstDiff~\cite{geng2023instructdiffusion} and MagicBrush~\cite{zhang2024magicbrush} tends to over-edit irrelevant regions (see toilet). SDEdit~\cite{meng2021sdedit} suffers from inferior image quality. Our method well balances the content preservation and instruction conformation.
  }
  \label{fig:qual}
\end{figure*}

\textbf{Image Generation Paradigm.} Since there are no similar studies supporting the direct instruction of fMRI signals, we adapt state-of-the-art image manipulation methods to evaluate the instruction capability of our model. Specifically, we first generate fMRI instructions using our approach. We then use the obtained intermediate reconstruction $Y^{Rec}$, along with the instruction text, as input to these methods for editing. 

\paragraph{Evaluation Results.} We demonstrate the qualitative evaluation results in Figure~\ref{fig:qual}. We observe that InstructPix2Pix~\cite{brooks2023instructpix2pix} tends to preserve its original content and struggles to follow instructions (see the last 4 rows). In contrast, InstructDiffusion~\cite{geng2023instructdiffusion} and MagicBrush~\cite{zhang2024magicbrush} are capable of better following instructions, but often change irrelevant attributes (e.g., the object shape such as the toilet and robot, or the background). For SDEdit~\cite{meng2021sdedit}, it also faces challenges in preserving unrelated regions and suffers from inferior image quality. 
In contrast, our approach strikes a balance between content preservation and instruction conformation (See the pose of the woman and the cactus shape), which we speculate is attributed to the dual-stream architecture where intermediate features are able to influence instruction progress constantly.
It is worth noting that these approaches not only rely on our intermediate reconstruction results but also require tedious sequential operations. In contrast, our proposed framework enables direct fMRI signal instruction, streamlining the process.

\subsection{Additional Results}



\begin{figure*}[t!]
  \centering
  \includegraphics[width=1.0\linewidth]{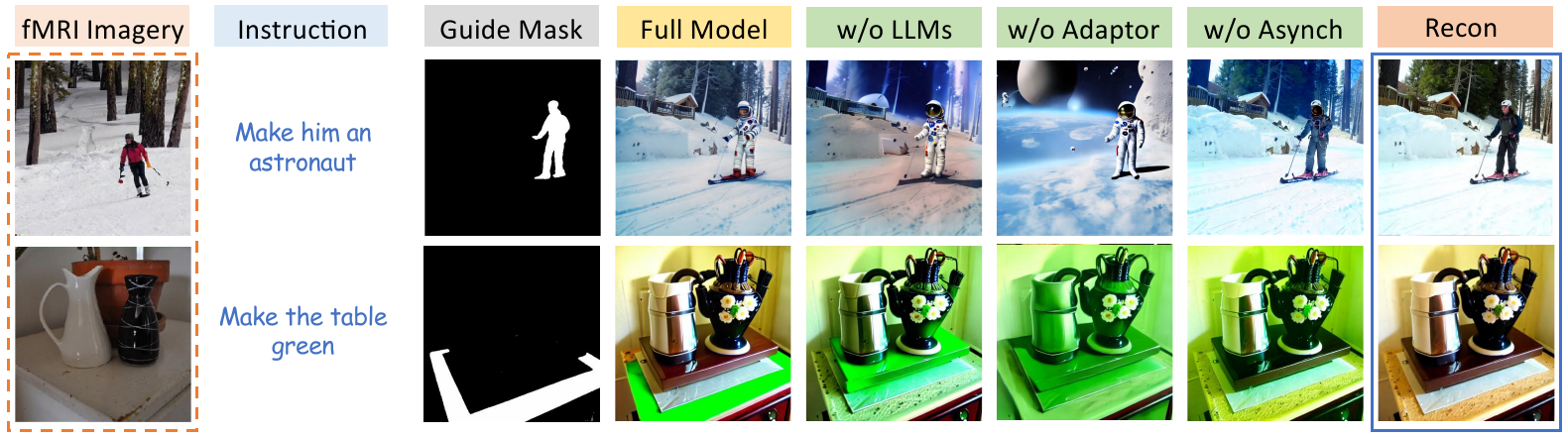}
  \caption{\textbf{Ablation Study.} Without feature injection from adaptor, the model struggles on balancing content preservation and instruction conformation. Removing asynchronous strategy brings difficulty on instruction conformation. Through the incorporation of LLMs guided region, our framework is able to precisely operate on relevant spatial locations (see the snow background and table region).
  }
  \label{fig:ablation}
\end{figure*}

\paragraph{Ablation Study.} We conducted ablation studies focusing on three important aspects of our method: the asynchronous diffusion strategy, the injection of adaptor features, and the use of LLM-guided region-aware attention. The experiments were carried out by (1) removing the asynchronous paradigm, (2) eliminating adaptor feature injection, and (3) excluding the LLMs agent. The numerical results are shown in Table~\ref{table:ablation}, and the visualization results are presented in Fig.~\ref{fig:ablation}.
Eliminating the asynchronous strategy makes it difficult for the instruction flow to comprehend aligned visual content, resulting in inferior performance.
When adaptor feature injection is removed, the instruction flow lacks sufficient visual information to identify instruction-relevant regions. Consequently, the instruction results deviate from the interpreted visual contents of the first stream, leading to poor performance in terms of both CLIP and Dino image similarity.
Without LLMs, the model struggles to perform precise operations within the intended spatial regions.
The full model effectively strikes a balance between identity preservation and instruction conformation, achieving visually pleasant results.

\paragraph{Multi-Modal Instruction.} 
\begin{figure*}[t!]
  \centering
  \includegraphics[width=1.0\linewidth]{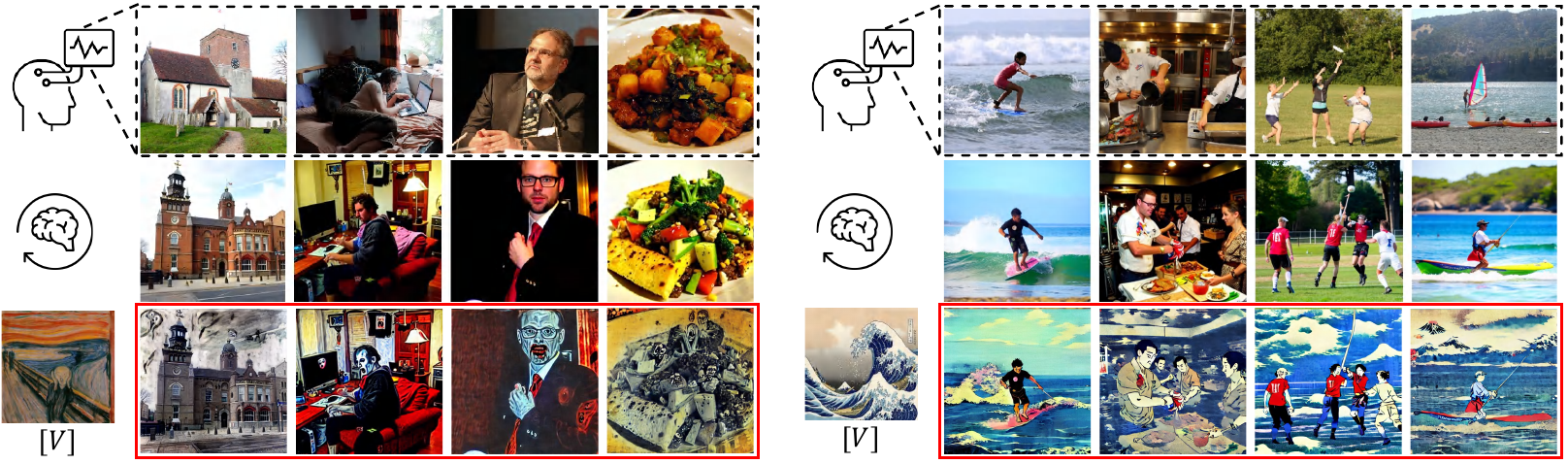}
  \caption{\textbf{Demonstration Results of Multi-Modal Instruction.} First row list the visual stimulus while second row depict our intermediate reconstructions. The manipulation results via \emph{In the [V] style} are shown within red boxes of the last row.}
  \label{fig:style}
\end{figure*}
Here we showcase the ability of our model to handle multi-modal instruction and achieve style manipulation. With minimal effort, our model can be extended to support visual instruction. Specifically, we modify the instruction description to ``In the [V] style'' where the style ``[V]'' can be obtained using image inversion approaches~\cite{ruiz2023dreambooth,mokady2023null}. It is evident that our approach effectively captures the style of the reference image and applies it to the target.

\section{Conclusion}

In this paper, we propose \textbf{\model{}}, a framework that enables users to effectively interact with human “dreams” and influence final presentations.
Our framework offers several appealing properties:
1) We \emph{pioneer a novel interaction paradigm with the human brain} by introducing a prototype that directly instructs fMRI signals via natural language descriptions, opening up exciting possibilities for future applications.
2) Our model's capability can be further enhanced using techniques that consider temporal and spatial perspectives, thereby improving the fidelity and precision of instructions.
3) With minimal effort, our system can seamlessly extend to support multi-modal instructions.

\paragraph{Limitations and Future Work.} 1) In this work, we consider the obtained biological brain signals from visual stimuli as representations of human ``thoughts''. However, in real-world applications, the situation is more complex. ``Dreams,'' for instance, might originate from internal brain activity rather than external stimuli. 2) Our model struggles with instructions involving the addition of small objects. 3) For the LLM-enhanced region-aware instructions, an extra forward process is required to first extract the instruction-relevant regions from the intermediate reconstructions. 4) Future work will explore other types of ``dreams,'' such as audio and text, and more complex interaction strategies, such as multi-turn conversations.


\paragraph{Ethical Issues.} As the title suggests, this model can be potentially exploited for malicious purposes such as mis-interpreting thoughts within human brain. We are committed to limiting the usage of our model strictly to research purposes.


\section{Declarations}

\paragraph{Availability of Data and Material.} 
Our model and the involved dataset will be publicly available. Meanwhile, we will also release the instruction construction pipeline. The used fMRI signal and instruction pairs will be released for the research community to further explore this task.

\paragraph{Competing Interests.} 
All authors certify that they have no affiliation or involvement in any organization or entity with any financial or non-financial interest in the subject matter or materials discussed in this manuscript.


\paragraph{Authors' Contributions.}
Yasheng Sun coordinated the teamwork and proposed the overall method. Bohan Li trained the fMRI-to-Image module and conducted comparison results.
Mingchen Zhuge initialized the idea of multi-modal learning with human brain activity signals and drafted the current version of the abstract and introduction. Prof. Fan manage the whole project. Prof. Fan, Prof. Salaman, Prof. Fahad, and Prof. Koike had insightful discussions and provided help polishing the manuscript. 
All authors read and approved the final manuscript.

\paragraph{Acknowledgements.} 
The authors express their gratitude to the anonymous reviewers and the editor, whose valuable feedback greatly improved the quality of this manuscript.

\bibliographystyle{unsrt}
\bibliography{egbib}

\newpage
\clearpage
\appendix
\appendixpage

\section{Implementation Details}
\paragraph{Optimization Objective.} 
In addition to the typical MSE Loss~\cite{ho2020denoising} used in the Latent Diffusion Model (LDM), we also incorporate StyleCLIP~\cite{patashnik2021styleclip} loss to guide the model toward the correct manipulation direction.  Formally,

\begin{equation}
\begin{aligned}
    \mathcal{L}_{style} &= 1 - \\
    \mathcal{D}(
    \textbf{F}_{vis}(\hat{Y}^{edit}) &- \textbf{F}_{vis}(Y), 
    \textbf{F}_{text}(T^{edit}) - \textbf{F}_{text}(T)),
\end{aligned}
\end{equation}

where $\mathcal{D}$ indicates the cosine similarity between two vectors. 
$Y$ and $T$ represent the original visual stimuli and its corresponding text, while the $\hat{Y}^{edit}$ and $T^{edit}$ are the synthesized result produced by our network and its corresponding ground truth caption, respectively. $\hat{Y}^{edit}$ can be estimated at each time step as follows:
\begin{equation}
\label{eq:decode_z}
    \hat{\textbf{z}}_0^e = (\textbf{z}_t^e - \sqrt{1-\overline{\alpha}_t} \bm{\epsilon}_{\theta} (\textbf{z}_t^e)) / \sqrt{\overline{\alpha}}_t,
\end{equation}
where the $\overline{\alpha}_t := \prod_{s=1}^t \alpha_s$. 
$\hat{Y}^{edit}$ can be decoded from $\hat{\textbf{z}}_0^e$.
The functions $\textbf{F}_{vis}$ and $\textbf{F}_{text}$ represent the CLIP backbone used to extract the visual and text features, respectively.
Such formulation ensures that the editing direction of the synthesized image aligns closely with the text direction in CLIP space.
The overall loss function, which combines the style direction loss with the foundational LDM loss, is mathematically represented as:
\begin{equation}
\mathcal{L}=\mathcal{L}_{simple}+\sqrt{\bar{\alpha}_t} \mathcal{L}_{style},
\end{equation}
where $\sqrt{\bar{\alpha}_t}$ is leveraged to emphasize the supervision with lower noise input (i.e., smaller time-step) and reduce the impact with higher noise (i.e., larger time-step).

\paragraph{Inference Paradigm.} Given a target fMRI signal $X$, our goal is to interact with it through the instruction $I$. The primary objective is to manipulate the visual content within the human brain toward the desired direction, resulting in $Y^{edit}$. During inference, we first use $\bm{g}_\theta(\textbf{z}^r_t, X, t)$ to perform $K$ diffusion steps for basic feature formation from $X$. Building on this, the dual-stream diffusion model $\bm{\epsilon}_\theta(\textbf{z}^r_t, \textbf{z}^e_t, I, X, t)$ progressively guides the fMRI signal.
After the diffusion process is complete, the latent code $\textbf{z}_0^e$ is transformed back to $Y^{edit}$ following Equation~\ref{eq:decode_z}. The detailed inference process is depicted in Algorithm~\ref{alg:test}.

\begin{algorithm*}[!ht]
  \caption{Inference of the asynchronous dual-stream diffusion model $\bm{\epsilon}_{\theta}$}
  \label{alg:sampling}
  \label{alg:test}
  \small
  \begin{algorithmic}[1]
    \vspace{.04in}
    \State $  \textbf{z}_T^e \sim \mathcal{N}(\mathbf{0}, \mathbf{I}), $ $  \textbf{z}_T^r \sim \mathcal{N}(\mathbf{0}, \mathbf{I})  $

    \For{$t=T, \dotsc, T-K+1$}
      \State $\textbf{z}^{nr} \sim \mathcal{N}(\mathbf{0}, \mathbf{I})$
      \State $\textbf{z}^r_{t-1} = \frac{1}{\sqrt{\alpha_t}}\left( \textbf{z}^r_{t} - \frac{1-\alpha_t}{\sqrt{1-\bar{\alpha}_t}} \bm{g}_\theta(\textbf{z}^r_{t}, X,t) \right) + \sigma_t \textbf{z}^{nr}$
    \EndFor

    \For{$t=T, \dotsc, K+1$}
      \State $\textbf{z}^{nr} \sim \mathcal{N}(\mathbf{0}, \mathbf{I})$ if $t > K+1$, else $\textbf{z}^{nr} = \mathbf{0}$
      \State $\textbf{z}^{ne} \sim \mathcal{N}(\mathbf{0}, \mathbf{I})$
      \State $\textbf{z}^r_{t\textcolor{red}{-K}-1} = \frac{1}{\sqrt{\alpha_{t\textcolor{red}{-K}}}}\left( \textbf{z}^r_{t\textcolor{red}{-K}} - \frac{1-\alpha_{t\textcolor{red}{-K}}}{\sqrt{1-\bar{\alpha}_{t\textcolor{red}{-K}}}} \bm{g}_\theta(\textbf{z}^r_{t\textcolor{red}{-K}}, X,t\textcolor{red}{-K}) \right) + \sigma_{t\textcolor{red}{-K}} \textbf{z}^{nr}$
      \State $\textbf{z}^e_{t-1} = \frac{1}{\sqrt{\alpha_t}}\left( \textbf{z}^e_{t} - \frac{1-\alpha_t}{\sqrt{1-\bar{\alpha}_t}} \bm{\epsilon}_\theta(\textbf{z}^r_{t\textcolor{red}{-K}}, \textbf{z}^e_{t}, I,X,t) \right) + \sigma_t \textbf{z}^{ne}$
    \EndFor

    \For{$t=K, \dotsc, 1$}
      \State $\textbf{z}^{ne} \sim \mathcal{N}(\mathbf{0}, \mathbf{I})$ if $t > 1$, else $\textbf{z}^{ne} = \mathbf{0}$
      \State $\textbf{z}^e_{t-1} = \frac{1}{\sqrt{\alpha_t}}\left( \textbf{z}^e_{t} - \frac{1-\alpha_t}{\sqrt{1-\bar{\alpha}_t}} \bm{\epsilon}_\theta(\textbf{z}^r_0, \textbf{z}^e_{t}, I,X,t) \right) + \sigma_t \textbf{z}^{ne}$

    \EndFor
    
    \State \textbf{return} $\textbf{z}^e_0$
    \vspace{.04in}
  \end{algorithmic}
\end{algorithm*}

\paragraph{Training Paradigm.}
During training, the denoising operation is performed asynchronously, with the instruction flow lagging behind by $K$ time steps. The detailed training process of our dual-stream diffusion framework is outlined in Algorithm~\ref{alg:training}.

\begin{algorithm*}[h]
  \caption{Training of the asynchronous dual-stream diffusion model $\bm{\epsilon}_{\theta}$}
  \label{alg:training}
  \small
  \begin{algorithmic}[1]
    \Repeat
      \State $(\textbf{z}_0^e, \textbf{z}_0^r, I, X) \sim q(\textbf{z}_0^e, \textbf{z}_0^r, I, X)$
      \State $\bm{\epsilon} \sim \mathcal{N}(\mathbf{0}, \mathbf{I}), \bm{\epsilon}^r \sim \mathcal{N}(\mathbf{0}, \mathbf{I})$
      \State $t \sim \text{Uniform}({K+1,\ldots,T})$
      \State Take a gradient descent step on
      \State 
      $ \quad \nabla_\theta \left\| \bm{\epsilon} - \bm{\epsilon}_\theta(\sqrt{\bar{\alpha}_{t\textcolor{red}{-K}}} \textbf{z}_0^r+\sqrt{1-\bar{\alpha}_{t\textcolor{red}{-K}}} \bm{\epsilon}^r, 
      \sqrt{\bar{\alpha}_{t}} \textbf{z}_0^e+\sqrt{1-\bar{\alpha}_{t}} \bm{\epsilon}, I, X, t) \right\|$
    \Until{converged}
  \end{algorithmic}
\end{algorithm*}

\section{More Ablation Studies}
In this section, we provide a detailed quantitative analysis of the components involved in our method.
Firstly, we present both qualitative and quantitative results to illustrate the impact of varying asynchronous steps.
Next, we offer further insights into the design of the dual-stream architecture.
Finally, we include additional qualitative ablation results to demonstrate the effect of each part of our architecture.

\paragraph{Impact of Varied Asynchronous Steps.} To investigate the impact of delayed steps on model performance, we conduct experiments with varied time steps. Specifically, in our implementation, we adopt $T=50$ at inference for efficiency. Thus, we choose $K=0, 5, 15, 25, 35$, which are broadly distributed, to conduct the experiment. For the sake of comparison, we do not incorporate LLMs and manually fix the random seed for all settings. Table~\ref{table:ablate_time_steps} illustrates the impact of delayed time steps. We observe that overall performance improves as we increase the delayed time steps, but the improvement becomes marginal beyond $K=15$. This is likely because $K=15$ is sufficient for basic feature formation.

\begin{table*}[h] 
\setlength{\tabcolsep}{28pt}
\begin{center}  
\caption{\textbf{The impact of different delayed time steps on model performance.}}
\label{table:ablate_time_steps}
\begin{tabular}{lcccccccc}
\toprule
Delay Steps & 0 & 5 &  15 & 25 &
35 \\

\midrule  
\textbf{CLIP-I} & 0.636 & 0.638 & \textbf{0.642} & 0.642 & 0.641  \\
\textbf{Dino-I} & \textbf{0.307} & 0.307 & 0.295 & 0.295 & 0.293 \\
\textbf{CLIP-D} & 0.108 & 0.109 & \textbf{0.114} & 0.106 & 0.108 \\
\bottomrule
\end{tabular}
\end{center}
\end{table*}

\begin{figure*}[ht!]
  \centering
  \includegraphics[width=1.0\linewidth]{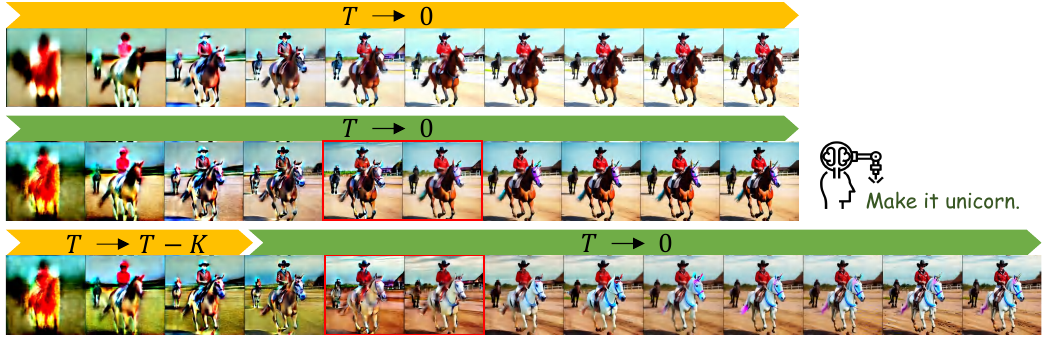}
  \caption{The first row displays the intermediate reconstruction results, while the second and third rows showcase the instruction performance using synchronous and asynchronous paradigms, respectively. The orange progress bar indicates the reconstruction timeline, while the green progress bar represents the instruction timeline.}
  \label{fig:async_demo}
\end{figure*}

\begin{figure*}[ht!]
  \centering
  \includegraphics[width=1.0\textwidth]{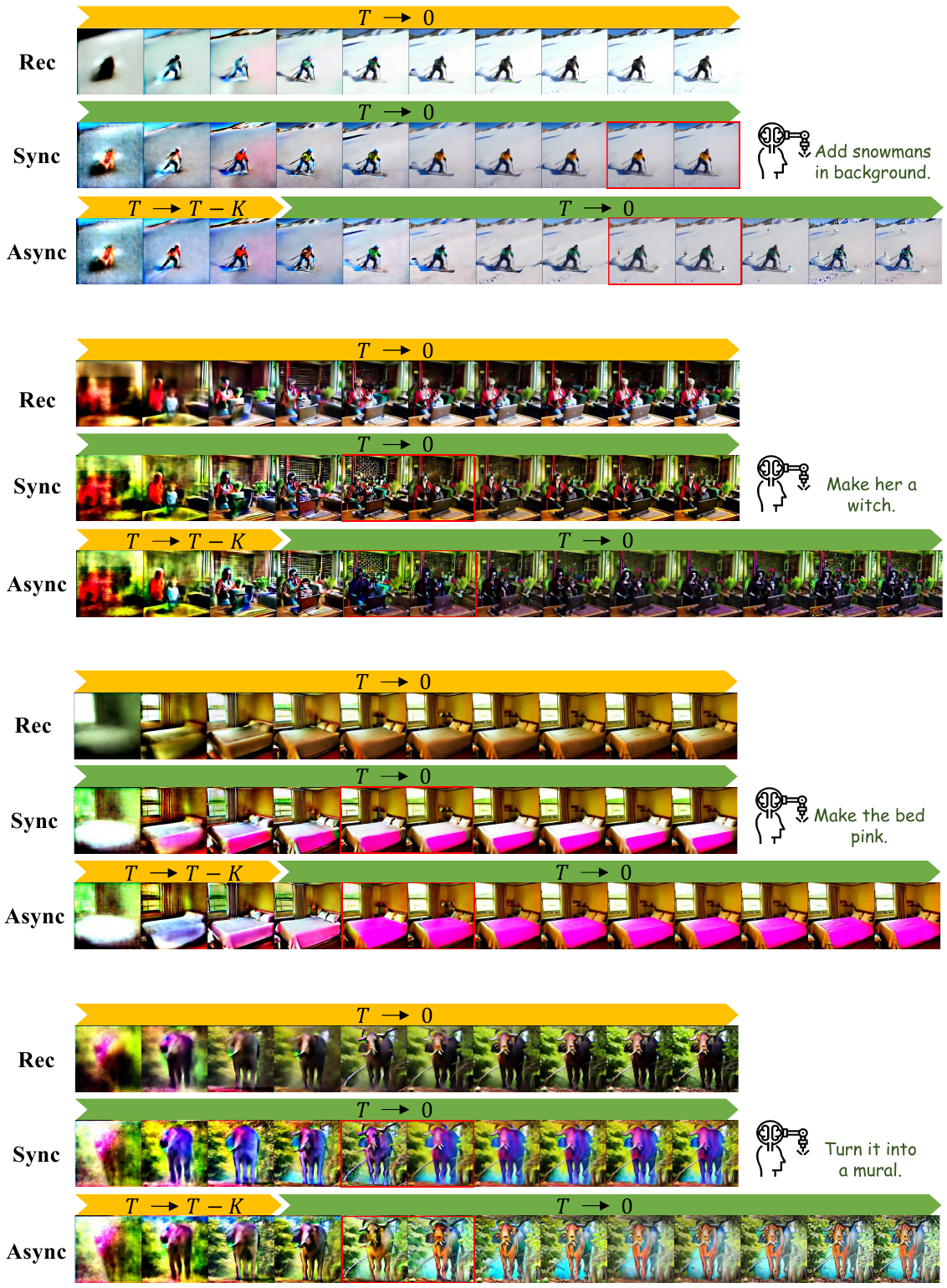} 
  \caption{The first row displays the intermediate reconstruction results, while the second and third rows showcase the instruction performance using synchronous and asynchronous paradigms, respectively. The orange progress bar indicates the reconstruction timeline, while the green progress bar represents the instruction timeline.}
  \label{fig:async_all_supp}
\end{figure*}

To further demonstrate the effect of the asynchronous strategy, we present the intermediate manipulation process in Fig.\ref{fig:async_demo}. We compare the synchronous strategy (set $K=0$ in the second row) with the asynchronous strategy (set $K=15$ in the third row). The individual intends to "Make it a unicorn" to manipulate brain activity. In the synchronous scenario, the model fails to capture the pertinent structure, i.e., the horse, for effective instruction (as indicated by the color of the horse within the red box). Without allowing basic feature formation in advance, the instruction model encounters difficulty in identifying the relevant structure. For additional visual results, readers can refer to Fig.\ref{fig:async_all_supp}.

\begin{figure*}[ht!]
  \centering
  \includegraphics[width=1.0\textwidth]{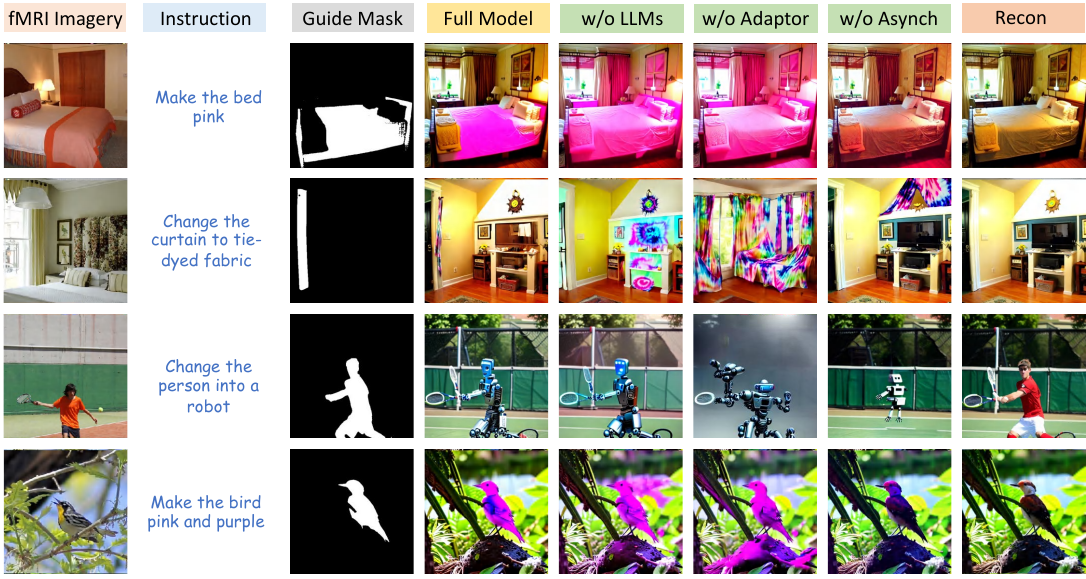} 
  \caption{Without feature injection from adaptor, the model struggles on balancing content preservation and instruction conformation. Removing asynchronous strategy brings difficulty on instruction conformation. Through the incorporation of LLMs guided region, our framework is able to precisely operate on relevant spatial locations.}
  \label{fig:more_ablation}
\end{figure*}

\paragraph{Impact of Injection Strategy.} The feature injection module acts as a connector, linking the instruction flow to the generation flow, ensuring seamless progressive bridging and effective instruction delivery. For the architecture design of our injection module, we explored both convolution-based structures~\cite{zhang2023adding} and transformer-based structures~\cite{hu2023animate}. Experiments demonstrate that the transformer-based architecture does not improve performance. We speculate this is because the convolution operation's advantage in maintaining spatial layout facilitates learning. Regarding the injection strategy, we empirically found that injection through ResBlock outperforms other blocks. Additionally, leveraging more advanced fusion techniques, such as transformers, does not enhance our performance.

\paragraph{More Qualitative Results.} We demonstrate additional visual results of our ablation study in Fig.~\ref{fig:more_ablation}. Eliminating the asynchronous strategy makes it difficult for the instruction flow to comprehend the aligned visual content, leading to inferior performance (see the bed and robot). Without the feature injection from the adaptor, the instruction flow lacks sufficient visual information to identify instruction-relevant regions (see the mis-edited curtain and bird). Without LLMs, the model struggles to perform precise operations within the intended spatial regions (see the bed and the bird). The full model effectively balances identity preservation and instruction conformation, achieving visually pleasing results.

\section{More fMRI Instruction Results}
In this section, we present additional visual instruction results. Although our approach is designed for language-guided instruction, it can readily extend to visual instruction with minimal effort.

\begin{figure*}[ht!]
  \centering
  \includegraphics[width=1.0\textwidth]{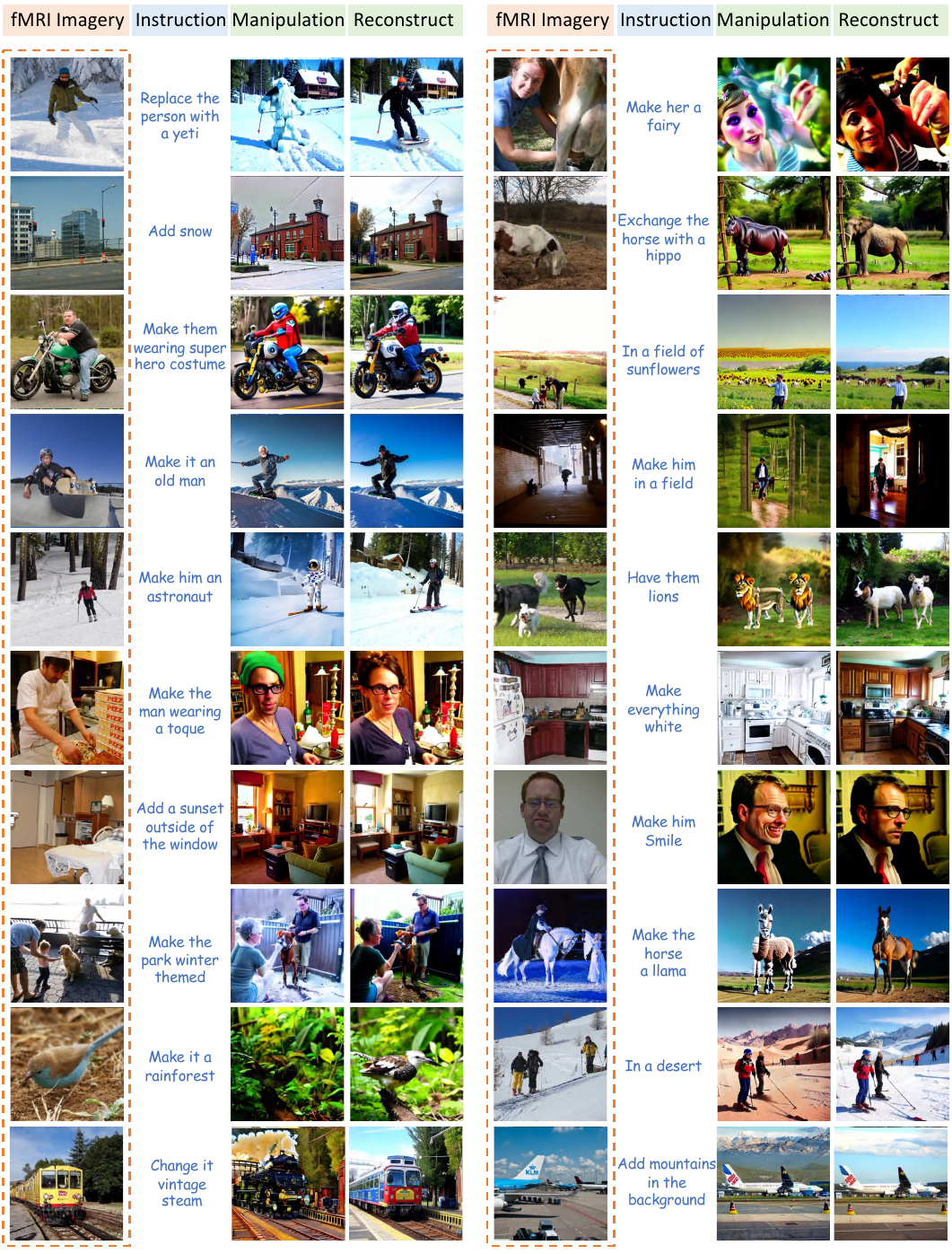} 
  \caption{\textbf{fMRI signal instruction with natural language description.} The first column displays the brain's visual stimulus. The second column illustrates the individual's intended operation. The third column presents the instruction results, while the fourth column shows the intermediate reconstruction results generated by our model.}
  \label{fig:supp_edit0}
\end{figure*}

\begin{figure*}[ht!]
  \centering
  \includegraphics[width=1.0\textwidth]{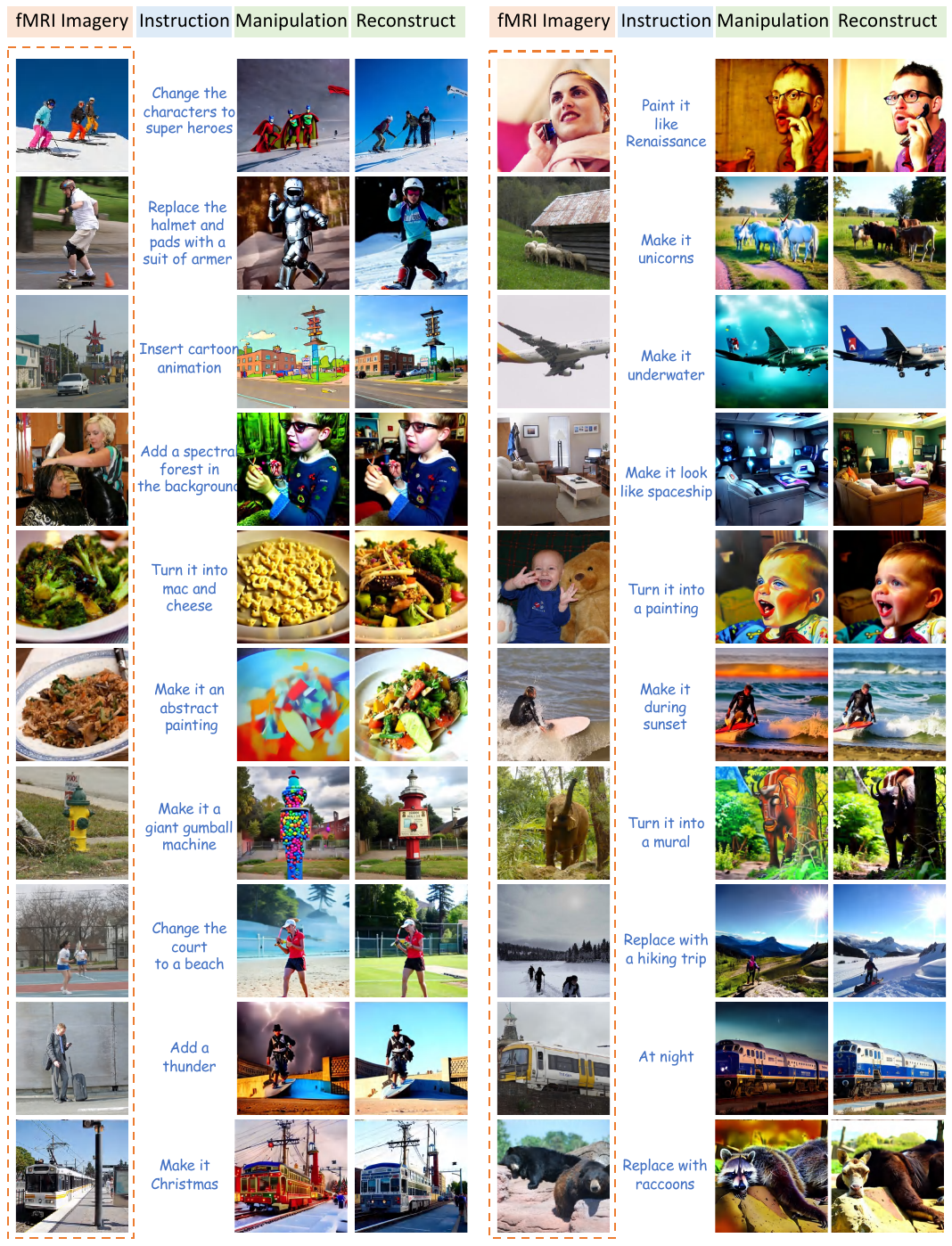} 
  \caption{\textbf{fMRI signal instruction with natural language description.} The first column displays the brain's visual stimulus. The second column illustrates the individual's intended operation. The third column presents the instruction results, while the fourth column shows the intermediate reconstruction results generated by our model.}
  \label{fig:supp_edit1}
\end{figure*}

\paragraph{Natural Language Instruction.}
We demonstrate more instruction results in Fig.~\ref{fig:supp_edit0} and Fig.~\ref{fig:supp_edit1}.
It can be seen that our approach is able to effectively interface with brain activity conforming to human instructions.

\begin{figure*}[ht!]
  \centering
  \includegraphics[width=1.0\textwidth]{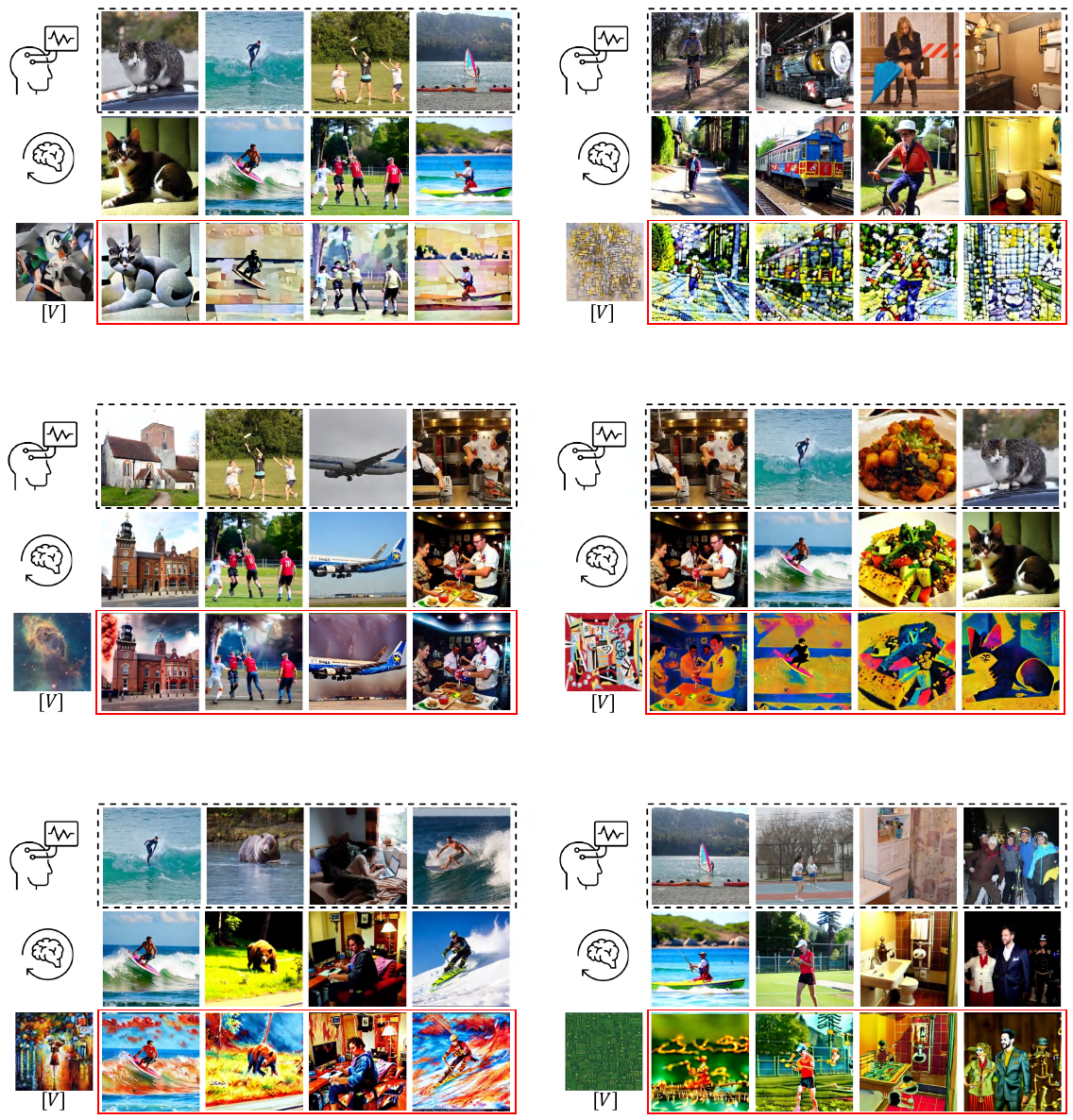} 
  \caption{\textbf{Demonstration Results of Multi-Modal Instruction.} First row list the visual stimulus while second row depict our intermediate reconstructions. The manipulation results via \emph{In the [V] style} are shown within red boxes of the last row.}
  \label{fig:supp_style}
\end{figure*}

\paragraph{Style Manipulation.}
For style manipulation, we first invert the style reference image to a word embedding ${[V]}$ in the text latent space, following the common inversion strategy~\cite{ruiz2023dreambooth,mokady2023null}. We then use \emph{In the style of} $[V]$ as our new instruction. The results of this instruction are depicted in Fig.~\ref{fig:supp_style}. It can be seen that our approach effectively captures the style of the reference image and applies it to the target.

\end{document}